# INEQUIVALENCE OF ENSEMBLES IN STATISTICAL MECHANICS

## R.P. VENKATARAMAN

1371, XIII Main, II Stage, I Phase, B.T.M.Layout,

BANGALORE-560076, INDIA

## ABSTRACT

For studying the thermodynamic properties of systems using statistical mechanics we propose an ensemble that lies in between the familiar canonical and microcanonical ensembles. From a comparative study of these ensembles we conclude that all these ensembles may not yield the same results even in the thermodynamic limit except at high temperatures. An investigation of the coupling between systems suggests that the state of thermodynamic equilibrium is a special case of statistical equilibrium. As a byproduct of this analysis we have obtained a general form for probability density function in an interval.

## I. INTRODUCTION

In studying the thermodynamic properties of systems using statistical mechanics, we restrict ourselves to a constant energy surface since we know that the energy of the system under consideration is constant. The basic assumption of ergodicity which helps us replace the time average of observables by the corresponding phase averages leads us naturally to a constant density on the energy surface given by

$$\rho = a \text{ constant if } (E - \Delta E) \leq E \leq (E + \Delta E) \text{ and } 0 \text{ otherwise} \qquad (1)$$

The ensemble of systems represented by this equation is called the microcanonical ensemble in which all the systems have the same energy $E$. For most physical systems the use of microcanonical ensemble turns out to be formidable. Besides this Gibbs[1] proposed the canonical ensemble. Starting from the postulates that 1) for statistical ensemble density $\rho$ has to be a stationary state of Liouville's equation

$$[\rho, H] = 0 \qquad (2)$$

and that 2)

$$\int d\Gamma \, \rho = 1 \qquad (3)$$

where $H$ is the Hamiltonian and $d\Gamma$ is the volume element in phase space he proposed

$$\rho = \exp[-\alpha - \beta H] \qquad (4)$$

as "the most simple form conceivable". ***Exp {-α}*** is the normalisation constant in the above equation. He went ahead to identify **β** as **1/(kT),** where **k** is the Boltzman constant and **T** is the temperature, from the structure of a Pfaff differential which is analogous to that of **dF** in thermodynamics. The ensemble of systems represented by Eq. (4) is called the canonical ensemble of Gibbs. It was only many years later that Jaynes[2] showed that these ensembles can be derived from information theory[3,4]. Starting from this theory we embark upon the intermediate ensemble given by,

$$\rho = \exp[-\alpha - \beta H - (\gamma/N)H^2] \qquad (5)$$

where **γ/N** is the LaGrange multiplier corresponding to an additional constraint that has been introduced as a better approximation to the microcanonical ensemble, the subject and purport of this exercise. In section II we outline the method of information theory and derive all possible forms of density thereof. We also obtain general forms of probability density function[5] and compare them with those obtained from information theory. Since our aim is to study the thermodynamic properties of systems we identify, in section III the coefficients **β** and **γ/N** of Eq. (5) in terms of thermodynamic quantities and at once realise the physical reasoning for introducing higher order terms of **H** in **ρ**. In section IV we derive the analogous thermodynamic quantities. From a comparative study of the mean energy and fluctuations in all these ensembles we conclude in section V that canonical and microcanonical ensembles may not yield the same results even in the thermodynamic limit except at high temperatures. The coupling between systems described by various ensembles is investigated in section VI and the conditions for statistical and thermodynamic equilibria discussed. We summarise the results in section VII.

## II. INFORMATION THEORY AND DENSITY FUNCTIONS

The method of information theory is but the method of finding the most probable or least biased density function involving LaGrange's multipliers. To find[5] the most probable distribution **ρ** subject to a given set of linearly independent constraints that fixes all the moments

$$<x^i> = c_i \qquad (6)$$

where $k_{i's}$ are constants and **i** takes all nonnegative integer values, assign a statistical entropy to **ρ** defined by

$$S_\rho = -\langle k \ln \rho \rangle \qquad (7)$$

In Eq. (6) the brackets denote the averaging done with respect to the density function and **k** in Eq.(7) is introduced for thermodynamic reasons. The least biased form of **ρ** is obtained by minimising information, which is nothing but negentropy, subject to the complete set of linearly independent constraints given by Eq. (6). For any small variation in **ρ** we have

$$\delta \int d\Gamma \, \rho \, [-k \ln \rho - k\Sigma_{i=0}^{\infty} \alpha_i \, x^i] = 0 \tag{8}$$

which yields

$$\rho(x) \, dx = \exp[-\Sigma_{i=0}^{\infty} \alpha_i x^i] \, dx \tag{9}$$

where $\alpha_i$'s are LaGrange multipliers to be determined from Eq. (6). If the constraints are only $n$ in number then this method gives the most probable least biassed form of density whose first $n$ moments are fixed. For instance if we do not know anything about the system except that the particle number and the energy are fixed we get on the constant energy surface

$$\rho = \exp[-\alpha] \tag{10}$$

which is the microcanonical ensemble. *Exp($\alpha$)* gives the possible number of states with a given energy. If on the other hand we want to define the ensemble in such a way that the average value of the energy of the ensemble alone is fixed we obtain

$$\rho = \exp[-\alpha - \beta H] \tag{11}$$

which is the canonical ensemble and

$$\rho = \exp[-\alpha - \beta H - (\gamma/N)H^2] \tag{12}$$

would correspond to that ensemble obtained by conserving $\langle H \rangle$ and $\langle H^2 \rangle$. This ensemble is a subset of Gibb's ensemble and it encompasses the microcanonical ensemble. By conserving higher and higher powers of $H$ we would approach microcanonical ensemble. Hence

$$\rho = \exp[-\Sigma_{i=0}^{\infty} \beta_i H^i] \tag{13}$$

represents the microcanonical ensemble density and representing as it does the entire class of physical systems it must be nothing but a general form of probability density function in an appropriate interval and by truncating it any intermediate ensemble could be realised.

For a probability density function of one variable Eq. (9) can be derived in yet another way[5]. Let $\rho(x)dx$ be a continuous probability density function in *[a,b]* that has no real or imaginary zeros or singularities. From Taylor expansion of *ln$\rho(x)$* about $x_0 \varepsilon [a,b]$ we get

$$\ln \rho(x) = \ln \rho(x_0) + \sum_{i=1}^{\infty} (x-x_0)^i 1/i! [\frac{d^i}{dx^i} \ln \rho(x)]_{(x=x_0)} \tag{14}$$

and from the above

$$\rho(x)dx = \exp[-\sum_{i=0}^{\infty} a_i x^i]dx \tag{15}$$

which is of the same form as Eq.(9) If $\rho(x)dx$ has real or imaginary zeros or singularities at *a* and/or *b* then the right hand side of Eq.(14) is not convergent and if we represent the zeros and singularities by *Z(x)* and *P(x)* respectively, Taylor expansion of *ln{ρ/(Z(x)P(x))}* still converges yielding

$$\rho(x)dx = Z(x)P(x)\exp[-\sum_{i=0}^{\infty} a_i x^i]dx \qquad (16)$$

Comparing Eq.(15) and Eq.(16) we realise that if information associated with the probability is redefined[5] as

$$I = \left\langle \ln\{\frac{\rho(x)}{Z(x)P(x)}\} \right\rangle \qquad (17)$$

then the results obtained from information theory and Taylor expansion tally. If *ρ(x)dx* has a finite number of real or imaginary zeros or singularities or a finite number of finite discontinuities in the interior of the interval we only have to divide the given interval into subintervals to follow the above procedure.

This can also be extended[5] to the case of many variables. For example a probability density function of two variables arising from

$$\left\langle x^i y^j \right\rangle = c_{ij} \qquad (18)$$

yields

$$\rho(x,y)dxdy = \exp[-\sum_{(i,j)=(0,0)}^{\infty} a_{ij} x^i y^j]dxdy \qquad (19)$$

where *(x,y)εD⊆R²* and *ρ* does not vanish anywhere in *D*. Coefficients *a_{ij}* are derived from

$$\left\langle \frac{-\partial}{\partial a_{ij}}\{\ln \rho(x)\} \right\rangle = c_{ij} \qquad (20)$$

Eq. (20) also can be derived[5] from Taylor expansion. Vide appendix.

### III. IDENTIFICATION OF LAGRANGE MULTIPLIERS

Having derived the formal expression for ensemble densities we now proceed to identify the LaGrange multipliers in terms of thermodynamic quantities. The transition from statistical mechanics to thermodynamics is effected in one step where the statistical entropy of the microcanonical ensemble is identified as the thermodynamic entropy. From Eq. (7) we get

$$S_\rho = k \ln g(E) \qquad (21)$$

where is *g(E)* is the possible number of states with a given energy *E*. From Eq.(21) the LaGrange multipliers occurring in Eq.(12) and Eq.(13) can be easily evaluated. It ensures, as will be shown below, that all the terms in the exponent in Eq. (13) are linear in *N*, the particle number. With this objective let us apply the microcanonical ensemble to the system and the rest of the system. Let $E_S$ and $E_R$ and $N_S$ and $N_R$ and $V_S$ and $V_R$ denote the energy, particle number and volume of the two respectively such that $E_S+E_R+E_{int} = E$ where $E_{int}$ is the interaction energy much smaller than $E_S$. The entropy of the rest of the system is

$$S_R(E_R) = S_R(E) + \sum_{i=1}^{\infty} (-E_S)^i / i! [\partial^i / \partial E^i S_R(E_R)]_{(E_R = E)} \quad (22)$$

by Taylor expansion and

$$\frac{\partial S_R}{\partial E_R} = \frac{1}{kT} \quad (23)$$

$$\frac{\partial^2 S_R}{\partial E_R^2} = \frac{-1}{kT^2 C_{V_R}} \quad (24)$$

where the temperature *T* and the specific heat $C_V$ refer to the rest of the system. Since the system and the rest of the system are decoupled- $E_{int}<<E_S$, the ensemble density for the system obeys the following equation.

$$\rho \propto g_R(E_R) \quad (25)$$

If $g_R(E_R)$ is large enough so that $E_R$ can be treated as a continuous variable we obtain from Eq. (21), Eq. (22), Eq. (23) and Eq. (24)

$$\rho = \exp[-E_S/kT - E_S^2/(2kT^2 C_V) - ....] \quad (26)$$

In the above equation whereas the first exponent varies as $N_S$ the second varies as $N_S^2/N_R$. In the limit when $(N_S/N_R) \rightarrow 0$ the second and higher order terms in $E_S$ vanish thus yielding Gibbs ensemble. This limit defines the concept of infinite reservoir. An infinite reservoir is the requisite of an experimentalist. When he measures the temperature of a system he should measure it in such a way that the system is itself is not in the least disturbed. Needless to say that it need not be and should not be introduced in the theoretical definition of equilibrium. For measuring temperature experimentally we do need it. It must be added that Gibbs[1] brings in the concept of heat bath only while interpreting *exp(-βH)* and not for defining the equilibrium state. Thus when we relax the condition $(N_S/N_R) \rightarrow 0$ the higher order terms in $E_S$ would no longer be negligible.

How do we decide what $(N_S/N_R)$ should be? When we say that the system is in equilibrium at a temperature *T,* then every macroscopic subsystem is in equilibrium with every other equally large macroscopic subsystem at the same temperature *T* which would also ensure equilibrium with infinite heat bath. But the latter does not imply the former. Hence we define the equilibrium of a system with an equally large macroscopic subsystem of identical nature and dropping the subscripts and introducing the normalisation factor we get

$$\rho = \exp[-\alpha - \beta H - (\gamma/N)H^2 ...] \tag{27}$$

$$\beta = \frac{1}{kT} \tag{28}$$

$$\gamma = \frac{-k\beta^2}{2C_I} \tag{29}$$

where $C_I$ is the input specific heat at constant volume per particle The ensemble density is first expressed as a function of $H$ and while integrating or summing over the appropriate states it is multiplied by $g(E)$. Hence Eq. (27) is expressed as a function of $H$ though the earlier equations in this section are all a function of $E$, referring as they do to microcanonical calculation. Eq. (27) represents the microcanonical ensemble and it can be shown from higher derivatives of $S_R$ that every term in the exponent is linear in $N$. No other definition of equilibrium i.e. $N_S/N_R \neq 1$ would have lead to the microcanonical density for the system. For a simple system like that of $N$ harmonic oscillators where the density of states is known one could verify that Eq.(27) indeed represents the reciprocal of density of states. By specifying the first few LaGrange multipliers we are in effect trying to approximate $1/g(E)$. The canonical ensemble has one input parameter $T$, while the inclusion of second order term in $H$ calls for another input parameter, $C_I$. The whole purpose of this exercise is to show through the intermediate ensemble

$$\rho = \exp[-\alpha - \beta H - (\gamma/N)H^2] \tag{30}$$

that the average thermodynamic properties obtained from all these ensembles need not be the same even in the thermodynamic limit except at high temperatures. The justification or the lack of it for neglecting higher order terms in $H$ would depend upon the temperature. From the form of the density it is easy to see that the higher order terms would have significant effect at low temperatures. In the next section we give a recipe for calculating the thermodynamic properties from all these ensembles.

### IV. EVALUATION OF THERMODYNAMIC QUANTITIES

Define the new partition function for the intermediate ensemble as

$$Z = \exp(\alpha) = \int d\Gamma \ \exp[-\beta H - (\gamma/N)H^2] \tag{31}$$

Then by Eq. (7)

$$S_\rho = k[\alpha + \beta \langle H \rangle + (\gamma/N)\langle H^2 \rangle] \tag{32}$$

From the above equations, replacing the average energy of the ensemble by $U$ we get

$$-kT \ln Z = U - T[S_\rho - k(\gamma/N)\langle H^2 \rangle] \tag{33}$$

For any other intermediate ensemble we have

$$-kT \ln Z = U - T[S_\rho - k \sum_{i=2}^{m} \beta_i \langle H^i \rangle] \tag{34}$$

Comparing Eq. (33) and Eq. (34) with the thermodynamic relation

$$F = U - TS \tag{35}$$

we define the free energy and the thermodynamic entropy as

$$F == -kT \ln Z \tag{36}$$

$$S = S_\rho - \sum_{i=2}^{m} \beta_i \langle H^i \rangle \tag{37}$$

the average energy of the ensemble being the natural choice for average thermodynamic energy. That *exp(-βF)* always represents the partition function could be easily interpreted as the maximisation of volume in phase space representing the minimisation of free energy in an appropriate way. Eq.(37) represents the only choice of *S* that would lead to thermodynamic equilibrium between two systems when $S_1+S_2$ is maximised. This will be discussed in section VI. For any *ρ* choose *S* such that this condition holds good. Thus in Eq.(36) and Eq.(37) no other regrouping of terms could be chosen to represent the thermodynamic entropy.

The other thermodynamic quantities cannot in general be expressed as derivatives of free energy. The reason lies in the following. The thermodynamic relation

$$TdS = dE + \sum_{i=1}^{n} X_i dx_i \tag{38}$$

where *$x_i$'s* including *E* and *S* are the state variables and the partial derivatives *$X_i$'s* are the conjugate forces is a micro law valid for every member in the ensemble with a given energy *E*. It arises from

$$T[(S+dS) - S] = [(E+dE) - E] + \sum_{i=1}^{n} X_i[(x_i+dx_i) - x_i] \tag{39}$$

Using Ehrenfest principles[6] if we want to calculate $X_i$ from the relation

$$X_i = -[\partial E / \partial x_i]_{(x_1, x_2, \ldots x_{(i-1)}, x_{(i+1)} \ldots x_n), S} \tag{40}$$

where the constancy of *S* refers to the constancy of the density of states at a given energy *E*, we have to bear in mind that systems with a given energy *E* occur with a definite probability. Hence, on multiplying Eq. (40) by *ρ* and integrating between appropriate limits, we get, in the energy representation

$$\langle X_i \rangle = \exp(-\alpha) \int_0^\infty d\Gamma - [\partial E / \partial x_i]_{(x_1, x_2, \ldots x_{(i-1)}, x_{(i+1)} \ldots x_n), S} \, g(E) \rho(E) \tag{41}$$

By multiplying Eq. (38) by the probability density and integrating between appropriate limits we obtain

$$\langle dE \rangle = \langle TdS \rangle - \sum_i \langle X_i dx_i \rangle \tag{42}$$

from which it is not possible to derive Eq. (41) and hence there is no way one could derive

$$d\langle E\rangle=(\partial\langle E\rangle/\partial\langle S\rangle)_{(V,N)}d\langle S\rangle-(\partial\langle E\rangle/\partial V)_{(S,N)}dV+(\partial\langle E\rangle/\partial N)_{(S,V)}dN \quad (43)$$

which is what is used in deriving the conjugate forces in the canonical ensemble formulation. It means that although in Eq. (39) the infinitesimals can be taken about the average values of state variables, with $S$ and $E$ replaced by their average values, the conjugate forces, which are given by the partial derivatives of the former have to be derived only from Eq. (41). So $\langle X_i\rangle$ need not in general satisfy the relation

$$X_i=[\partial U/\partial x_i]_{(x_1,\ldots S)} \quad (44)$$

though it may be valid for certain cases like the canonical ensemble. For example in the case of the intermediate ensemble pressure is given by

$$P=\int_0^\infty dE-[\partial E/\partial V]_{(N,S)}\ g(E)\exp(-\alpha-\beta E-(\gamma/N)E^2) \quad (45)$$

and is not equal to

$$\partial/\partial V[-kT\ln Z] \quad (46)$$

Eq. (45) alone ensures the validity of virial theorem: **PV=2/3U** for free particles which can be derived from mechanical considerations alone and Eq. (44) violates it. In short the processes of ensemble averaging and Legendre transformation need not commute. Ensemble average has to follow Legendre transformation and not precede it. The recipe hence is to apply Eq. (38) to every microcanonical ensemble and use Eq. (36), Eq. (37) and Eq. (41) for evaluating the thermodynamic properties.

    To complete the analogy between thermodynamics and statistical mechanics one has to show that the thermodynamic limit exists. With the intermediate ensemble for a hard core finite range attractive potential the existence of the thermodynamic limit can be proved along the lines discussed in Thopmson[7] for Gibbs ensemble. In all these derivations the laws of thermodynamics are kept in tact. The transition from statistical mechanics to thermodynamics is made through the axiom given by Eq. (21) and whatever we derive with different densities has to be and is consistent with thermodynamics. These calculations therefore do not represent a rederivation of the macroscopic thermodynamic laws from microscopic mechanics.

## V. AVERAGE ENERGY AND FLUCTUATIONS

In the energy representation, partition function for Eq.(30) can be represented by

$$Z=\int_0^\infty dE\ g(E)\exp[-\beta E-(\gamma/N)E^2] \quad (47)$$

Taylor expansion of *lng(E)* yields,

$$Z = g(\bar{E})[\int_0^\infty dE \exp\{-\beta(E-\bar{E})-(\gamma/N)(E-\bar{E})^2$$

$$+ \sum_{i=1}^\infty \frac{(E-\bar{E})^i}{i!} \{\frac{\partial^i}{\partial E^i}\ln g(E)\}_{(E=\bar{E})}\}]$$

(48)

It is obvious that the integrand does not always have a saddle point as it depends on the signature of the coefficients. In evaluating the canonical partition function

$$Z = \int_0^\infty dE\, g(E)\exp(-\beta E) \tag{49}$$

(50)

is claimed to be the saddle point equation. It is not the saddle point equation but the defining equation for $\bar{E}(T)$ in the microcanonical ensemble calculation. Hence

$$F = E - Tk \ln g(E) \tag{51}$$

is the relation between averages obtained from microcanonical ensemble. It is important to note that *F* is not given by the partition function in by Eq. (49). Comparing the partition functions given by Eq. (48) and Eq. (49) we infer readily that for a given *T* average energy in the intermediate ensemble is smaller than that in the canonical ensemble and as we conserve higher and higher powers of *H* in the ensemble density, the value of $\bar{E}$ decreases steadily and we get the lowest value in the microcanonical case. From the nature of the higher order multipliers which are all *T* dependent, we also realise that the change in $\bar{E}(T)$ is not one of mere scaling or shifting. Since the LaGrange multipliers are all inversely proportional to *T*, we conclude that when at high temperatures all these ensembles must yield the same results they may not yield the same results at low temperatures even in the thermodynamic limit, for, all the terms in the exponent area linear in *N*. This analysis does not exclude the possibility of certain systems giving the same results in microcanonical and canonical ensembles, like for instance classical ideal gas. In this case the density of states obeys a power law and we also know that both the canonical and microcanonical ensembles yield the same results. In such a case as this, the limiting ensembles yield the same results because the saddle point exists in the partition function. That is, by solving the canonical partition function by the saddle method we are in effect doing a microcanonical ensemble calculation. In other cases through the Hamiltonian and LaGrange multipliers we are in effect trying to approximate the density of states. In some cases there may not be a saddle point in the partition function but the average energy might have as in

$$g(E) = \frac{c}{[1+E]^{0.5}} \tag{52}$$

where *c* is a constant, though this might not represent any realistic case. Even in those cases where there is no saddle in the expression for average energy, the statistical entropy can always be solved by saddle point approximation for a suitable upper limit **m** in the truncated Eq. (13). In this case a microcanonical calculation would lead to a negative temperature as Eq. (50) does. But the canonical ensemble yields a relation anlogous to Eq. (35)!

That the average energy, entropy and free energy are related by a Legendre transformation - except for a difference due to terms of order **lnN** negligible in the thermodynamic limit is usually interpreted as the equivalence[8] of ensembles in the thermodynamic limit. We have shown here that given any density of states we could always find these thermodynamic quantities satisfying such a relation, from the axiom given by Eq. (21) and that these averages $\overline{X_i(T)}$ vary from ensemble to ensemble. Thus this relation is necessary but not sufficient. As the densities differ, the partition function, and hence the average energy and free energy differ and hence does thermodynamic entropy and the difference is more than in terms of order **lnN**.

It is easy to show that $\dfrac{\overline{E^2} - \overline{E}^2}{\overline{E}^2}$ goes inversely as *N* in all these ensembles. The second central moment decreases steadily as we move from canonical to microcanonical ensemble. In the case of the intermediate ensemble we have

$$\langle (E-\overline{E})^2 \rangle = \langle (E-\overline{E}_c)^2 \rangle - (\overline{E} - \overline{E}_c)^2 \tag{53}$$

where $\overline{E}_c$ is the average energy in the canonical ensemble and within the brackets representing the averaging with respect to the intermediate ensemble it remains a constant. Hence the second central moment in the intermediate ensemble is smaller than that in the canonical ensemble. The width decreases with the inclusion of higher powers of *H*. The specific heat per particle *C* obtained from the intermediate ensemble includes correction terms besides $C_I$. It is given by

$$kT^2 NC_R = \langle H^2 \rangle - \langle H \rangle^2 - kT^2 d/dT[\gamma/N\{\langle H^3 \rangle - \langle H \rangle \langle H^2 \rangle\}] \tag{54}$$

The grand canonical ensemble yields the same equation for $\langle X_i(T)/N \rangle$ as the original ensemble. But if we want to obtain $\langle N(T) \rangle$ it is important to include, for reasons similar to those mentioned with regard to the variable *E*, higher order terms in *N* and also cross terms involving *E* and *N*.

# VI. COUPLING BETWEEN SYSTEMS

Consider two systems of energy $E_1$ and $E_2$ and particle number $N_1$ and $N_2$ respectively. If the two systems are coupled to form a composite system isolated from the surroundings, then the composite system would be in equilibrium and the sum of the statistical entropies is a maximum. In general for any variation in the energies of a composite system of $n$ sub-systems

$$\delta[\sum_{i=1}^{n} S_{\rho i}] = 0 \qquad (55)$$

Since the total energy is fixed

$$\sum_{i=1}^{n} \delta E_i = 0 \qquad (56)$$

Condition in Eq. (55) corresponds to allowing maximum number of states for the composite system. If we use the intermediate ensemble to describe the systems, using Eq. (32) and Eq. (56) we could reduce this condition for the case $n=2$ to

$$(\beta_1 - \beta_2) + 2[(\gamma_1/N_1)\overline{E}_1 - (\gamma_2/N_2)\overline{E}_2] = 0 \qquad (57)$$

This condition for statistical equilibrium does not imply thermodynamic equilibrium. If the two systems are of the same nature, then, at thermodynamic equilibrium, it further demands that

$$\frac{\overline{E}_1}{N_1} - \frac{\overline{E}_2}{N_2} = 0 \qquad (58)$$

This, nothing but the extensive property of energy, will be true by very definition of equilibrium. If the two systems are described by microcanonical or canonical ensemble the condition for statistical equilibrium is the same as that for thermodynamic equilibrium. But if the composite system is made up of three or more systems none of these ensembles stipulates thermodynamic equilibrium as the condition for statistical equilibrium as can be readily inferred from Eq. (55) and Eq. (56). These would be valid in general because of the maximisation of $S_\rho$ and the nature of the condition would depend on what ensemble we use to describe the system. It is easy to verify that statistical equilibrium is a transitive property. Physically Eq. (54) means that if two systems at different temperatures were brought together the composite system when shielded from the surroundings would be in statistical equilibrium even before thermodynamic equilibrium is attained. Thus thermodynamic equilibrium is a special case of statistical equilibrium.

Finally at statistical equilibrium we should be able to cast the density in the same form as $\rho$. This means that we have to find an effective, and not equilibrium $\beta$ and $\gamma/N$ in terms of $\beta_i$ and $\gamma_i$. Starting with the expression for total statistical entropy and using the relation $\langle E_i^2 \rangle = \langle E_i \rangle^2$ on the assumption that the integral for the saddle point could be solved by saddle point approximation, we get

$$S_\rho = S_{\rho_1} + S_{\rho_2} = k[\alpha_1 + \beta_1 \overline{E}_1 + (\gamma_1/N_1)\overline{E}_1^2 + \alpha_2 + \beta_2 \overline{E}_2 + (\gamma_2/N_2)\overline{E}_2^2]$$
(59)

and using the condition for statistical equilibrium given in Eq. (55) we cast $S_\rho$ into the same form as $S_{\rho_i}$

$$S_\rho = k[\alpha + \frac{1}{2}(\beta_1 + \beta_2)\overline{E} + \frac{1}{4}(\frac{\gamma_1}{N_1} + \frac{\gamma_2}{N_2})\overline{E}^2]$$
(60)

where we have neglected $(\overline{E}_1 - \overline{E}_2)^2$ in comparison with $(\overline{E}_1 + \overline{E}_2)^2$. The effective values of $\beta$ and $(\gamma/N)$ are given by the coefficients in Eq. (60) and

$$\rho = \exp[-\alpha - \frac{\beta_1 + \beta_2}{2}H - \{\frac{\gamma_1}{4N_1} + \frac{\gamma_2}{4N_2}\}H^2]$$
(61)

is a stationary state of Liouville's equation. Eq. (61) represents the state of the system at $t=t_0$ when the individual systems are brought together and is the initial state of the composite system whose time evolution is dictated by Liouville's equation. The final equilibrium state of the composite system will also be given by Eq. (60) except that $\beta_1$ and $\beta_2$ would both correspond to the temperatures at which the two systems equilibrate.

## VII. CONCLUSION

Through a comparative study of ensembles ranging from canonical to microcanonical, we have proved that all these ensembles may not yield the same results even in the thermodynamic limit except at high temperatures. While a simple system like that of a system of harmonic oscillators may be investigated to verify these results and check the qualitative and quantitative differences, it must be pointed out that it is only an academic exercise. The more interesting cases are those where we do not know the density of states and try to approximate it through experimentally measured parameters. As a byproduct of this analysis we have obtained a general form for probability density function in an interval. Power series expansion for functions of several variables is derived and convergence questions addressed using spherical polar coordinates.

## Acknowledgements

The author acknowledges fruitful discussions with Drs.P.N.Guzdar, B.R.Sitaram, and S.Madan and the financial support extended by the Council of Scientific and Industrial Research during her stay at I.I.T. Kanpur and I.I.T. Bombay and the support extended by Physical Research Laboratory Ahmedabad and N.I.I.T.,Tiruvanamalai for permitting her to type the text.

## REFERENCES

1) J.W. Gibbs, Elementary Principles In Statistical Mechanics, Chapter4, Yale University Press, New Haven, Reprinted by Dover Publications, New York, 1960.

2) E.T. Jaynes, Phy. Rev. 106 pp 620-630 1957.

3) C.E. Shannon and W.Weaver, Mathematical Theory of Communications, University of Illinois Press, 1963.

4) D.S. Jones, Elementary Information Theory, Oxford University Press, 1979.

5) R.P.Venkataraman, "Is Schroedinger Equation Consistent with Information Theory?" in Proceedings of the International Conference on Mathematical Modelling of Nonlinear Systems, Kharagpur, pp 366-372 1999 and the references therein.

6) G.S. Wannier, Statistical Physics (Chapter 5) John Wiley and Sons, New York, 1966.

7) C.J. Thompson, Mathematical Statistical Mechanics, Chapter 3, The MacMillan Co., New York, 1972.

8) R.B.Griffiths, Jl. Math. Phys, 106 PP 1447-1451, 1965 and other standard text books in statistical mechanics.

9) R.Courant & F.John Introduction to Calculus and Analysis Volumes 1 and 2, Springer, New York, 1989.

10) Philip C.Curtis Jr., Multivariate Calculus with Linear Algebra, John Wiley and Sons, 1972.

11) R.P.Venkataraman, "Necessary and Sufficient Conditions for Differentiability of a Function of Several Variables"-Communicated to American Mathematical Monthly.

,.

## APPENDIX

A probability density of the form given in Eq. (19) can be used when solutions to differential equations are sought for. Before deriving it in yet another way power series in several real variables is considered below. It is well known[9] that starting from

$$P_n(x) = \sum_{i=0}^{n} a_i x^i \tag{62}$$

where

$$a_i = 1/i! [d^i/dx^i P(x)]_{(x=0)} \tag{63}$$

Taylor and his pupils defined power series starting from

$$f(x) = \sum_{i=0}^{N} x^i / i! [d^i / dx^i f(x)]_{(x=0)} + R_N \quad (64)$$

by taking the upper limit to infinity. Analogously if one begins from[10]

$$P_N(x, y) = \sum_{i+j=0}^{N} a_{ij} x^i y^j \quad (65)$$

and defines

$$a_{ij} = \frac{1}{i! j!} [\partial^{(i+j)} / \partial x^i \partial y^j P]_{(x=0, y=0)} \quad (66)$$

one could show for *a* and *k* real, since

$$(1 + ax)^k = 1 + k(ax)/1! + k(k-1)(ax)^2/2! + \ldots \quad (67)$$

is convergent for $|ax| < 1$,

$$(1 + xy)^k = 1 + k(xy)/1! + k(k-1)(xy)^2/2! + \ldots \quad (68)$$

is convergent for $|xy|<1$ or equivalently $r<1$ in polar coordinates, where $(x,y) \varepsilon R^2$. Similarly since the power series for $exp(a\ x)$ is convergent for all $x \varepsilon R$.

$$\exp(xy) = 1 + (xy)/1! + (xy)^2/2! + \ldots + (xy)^r/r! + \ldots \quad (69)$$

and the above series is convergent for all $(x,y) \varepsilon R^2$. Taylor expansion for two variables is defined by

$$f(x, y) = \sum_{(i,j)=(0,0)}^{N} a_{ij} x^i y^j + R_N \quad (70)$$

and the coefficients in (68) and (69) and (70) are precisely those derived from

$$a_{ij} = \frac{1}{i! j!} [\partial^{(i+j)} / \partial x^i \partial y^j f]_{(x=0, y=0)} \quad (71)$$

Hence power series in two variables can be defined as

$$f(x, y) = \sum_{(i,j)=(0,0)}^{\infty} a_{ij} x^i y^j \quad (72)$$

if convergence questions for functions of several variables as in Eq.(72) are addressed using polar coordinates. Equation (72) is the limiting form of Eq.(70) and Eq. (71) gives a more natural expression for $a_{ij}$ 's. Similarly the power series expression for logarithm could also be derived to arrive at Eq. (19). Extension to several variables is straightforward. Extrema of functions of several variables could also be found easily solving just one equation in the radial coordinate and checking the signature of the second derivative in that coordinate[11].